\RequirePackage{fix-cm}
\documentclass[smallextended]{svjour3}       
\smartqed  
\usepackage{graphicx}
\journalname{Preprint}
\begin{document}

\title{On the self-replicating properties of Riemann zeta zeros: A statistical study
}

\titlerunning{Self-replicating zeta zeros}        

\author{Jouni J. Takalo
}


\institute{Space physics and astronomy research unit,\\University of Oulu, POB 3000, FIN-90014, Oulu, Finland \\
              Tel.: +358-40-7172358\\
              \email{jouni.j.takalo@oulu.fi;jojuta@gmail.com} 
												ORCID ID 0000-0002-1807-6674  \\
}

\maketitle

\newpage

\begin{abstract}
We study distributions of differences of unscaled Riemann zeta zeros, $\gamma-\gamma^{'}$, at large distances. We show, that independently of the height, a subset of finite number of successive zeros knows the locations of lower level zeros. The information contained in the subset of zeros is inversely proportional to $ln(\gamma/(2\pi))$, where $\gamma$ is the average zeta of the subset. Because the mean difference of the zeros also decreases as inversely proportional to $ln(\gamma/(2\pi))$, each equally long segment of the line $\Re(z)=1/2$ contains equal amount of information. 
The distributions of differences are skewed towards the nearest zeta zero, or at least, in the case of very nearby zeros, the skewness always decreases when zeta zero is crossed in increasing direction. We also show that the variance of distributions has local maximum or, at least, a turning point at every zeta zero, i.e., local minimum of the second derivative of the variance. In addition, it seems that the higher the zeros the more compactly the distributions of the differences are located in the skewness-kurtosis -plane. The flexibility of the Johnson distribution allows us to fit the distributions nicely, despite of the values of skewness and kurtosis of the distributions.
\end{abstract}

\section{Introduction}
\label{intro}

It is well-known that the correlation between close pairs of nontrivial zeros of the Riemann zeta function (scaled to have unit average spacing) is \cite{Montgomery}

\begin{equation}
\label{eq:CUE}
	R_{2}\left(x\right) = 1\!-\!\left(\frac{sin(\pi\,x)}{\pi\,x}\right)^{2} .
\end{equation}

\begin{figure}
	\centering
	\includegraphics[width=0.8\textwidth]{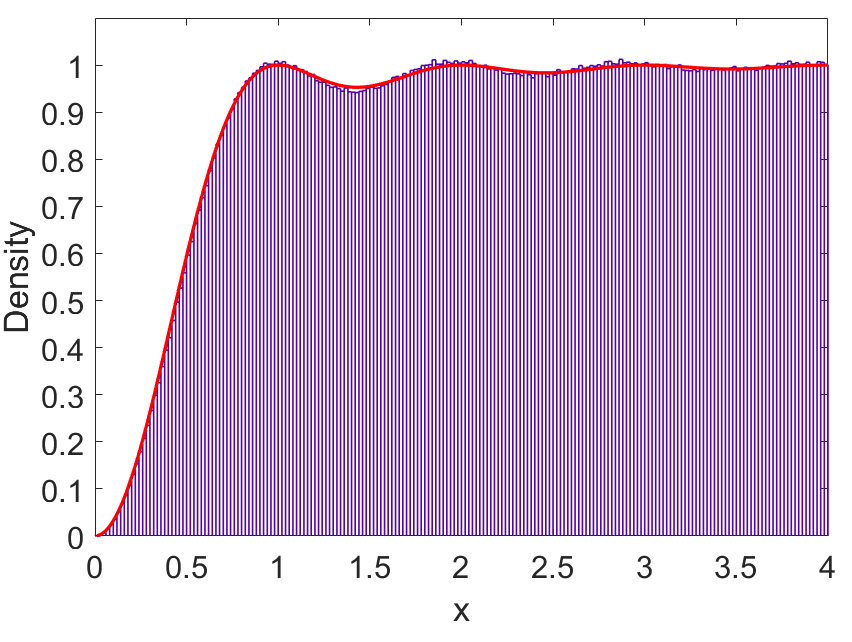}
		\caption{Distribution of locally close scaled differences of zeta zeros. Five million zeros tarting from billionth zero were used in the statistics. Red curve is the plot of function in Eq. \ref{eq:CUE}}.
		\label{fig:CUE_prediction}
\end{figure}

Figure \ref{fig:CUE_prediction} shows the density of 5 million differences starting from billionth zero and the prediction of Eq.\ref{eq:CUE} as a red curve. Odlyzko (1987) already showed that this conjecture was supported also by larger heights of zeta zeros \cite{Odlyzko}. The aforementioned papers are restricted to analyse only consecutive or locally close differences of zeros.  P\'erez-Marco (2011) has shown that the statistics of very large zeros do find the location the first Riemann zeros \cite{Perez-Marco}.
There have been several studies of the differences (pair correlations, triple correlations, n-point correlations) between zeta zeros. They predict troughs in the correlation functions at the sites of the zeros themselves \cite{Bogomolny_a}, \cite{Bogomolny_b}, \cite{Conrey}, \cite{Snaith}, \cite{Perez-Marco}, \cite{Rogers}, \cite{Takalo}. 
We study here distributions of differences of unscaled zeta zeros at longer distances, and analyse in more detail the reason for the troughs in correlations of zeta zeros using statistical methods.

\section{Data and methods}

\subsection{Nontrivial zeros of Riemann zeta function}

The data, i.e., the imaginary parts of nontrivial Riemann zeta zeros were fetched from  (https://www.lmfdb.org/zeros/zeta/) for the five million zero starting from 1st zero, 10 millionth zero, one billionth, 10 billionth and 100 billionth zero. The zeros starting from zero number $10^{23}$ were kindly provided by Dr. A. Odlyzko \cite{Hiary}.

\subsection{Johnson distribution}

Johnson distribution for the variable x is defined as 
\begin{equation}
z = \lambda+\delta\,ln\left(f\left(u\right)\right),
\end{equation}
with 
\begin{equation}
u = \left(x-\xi\right)/\lambda,
\end{equation}
Here z is a standardized normal variable and $f\left(u\right)$ has three different forms
the lognormal distribution, $S_{L}$:
\begin{equation}
f\left(u\right)=u,
\end{equation}
the unbounded distribution, $S_{U}$:
\begin{equation}
f\left(u\right)=u+{\left(1+u^{2}\right)}^{1/2},
\end{equation}
and the bounded distribution, $S_{B}$:
\begin{equation}
f\left(u\right)=u/\left(1-u\right).
\end{equation}
The supports for the distributions are $S_{L}: \xi<x,\; S_{U}: - \infty<x< \infty\; $and $S_{B}: \xi<x<\xi+\lambda$ \cite{Johnson}, \cite{Wheeler}. With these definitions, the probability distributions are for
$S_{L}$:
\begin{equation}
P\left(u\right)=\frac{\delta}{\sqrt{2\pi}}\,\times\,\frac{1}{u}\,\times\,exp\left\{-\frac{1}{2}\left[\gamma+\delta\ln\left(u\right)\right]^{2}\right\}.
\end{equation}
for $S_{U}$
\begin{equation}
P\left(u\right)=\frac{\delta}{\sqrt{2\pi}}\,\times\,\frac{1}{\sqrt{u^{2}+1}}\,\times\,exp\left\{-\frac{1}{2}\left[\gamma+\delta\ln\left(u+\sqrt{u^{2}+1}\right)\right]^{2}\right\}.
\end{equation}
and for $S_{B}$
\begin{equation}
P\left(u\right)=\frac{\delta}{\sqrt{2\pi}}\,\times\,\frac{1}{u/\left(1-u\right)}\,\times\,exp\left\{-\frac{1}{2}\left[\gamma+\delta\ln\left(\frac{u}{1-u}\right)\right]^{2}\right\}
\end{equation}

\section{Distributions of differences of zeta zeros}

\subsection{Variances and kurtoses of the distributions}

In the next we study the differences, $(\delta)$, of unscaled zeta zeros. We use the following notation

\begin{equation}
\label{eq:delta}
\delta(n) = \gamma(n+i)-\gamma(i), n=1,2,3,...
\end{equation}

\begin{figure}
	\centering
	\includegraphics[width=0.8\textwidth]{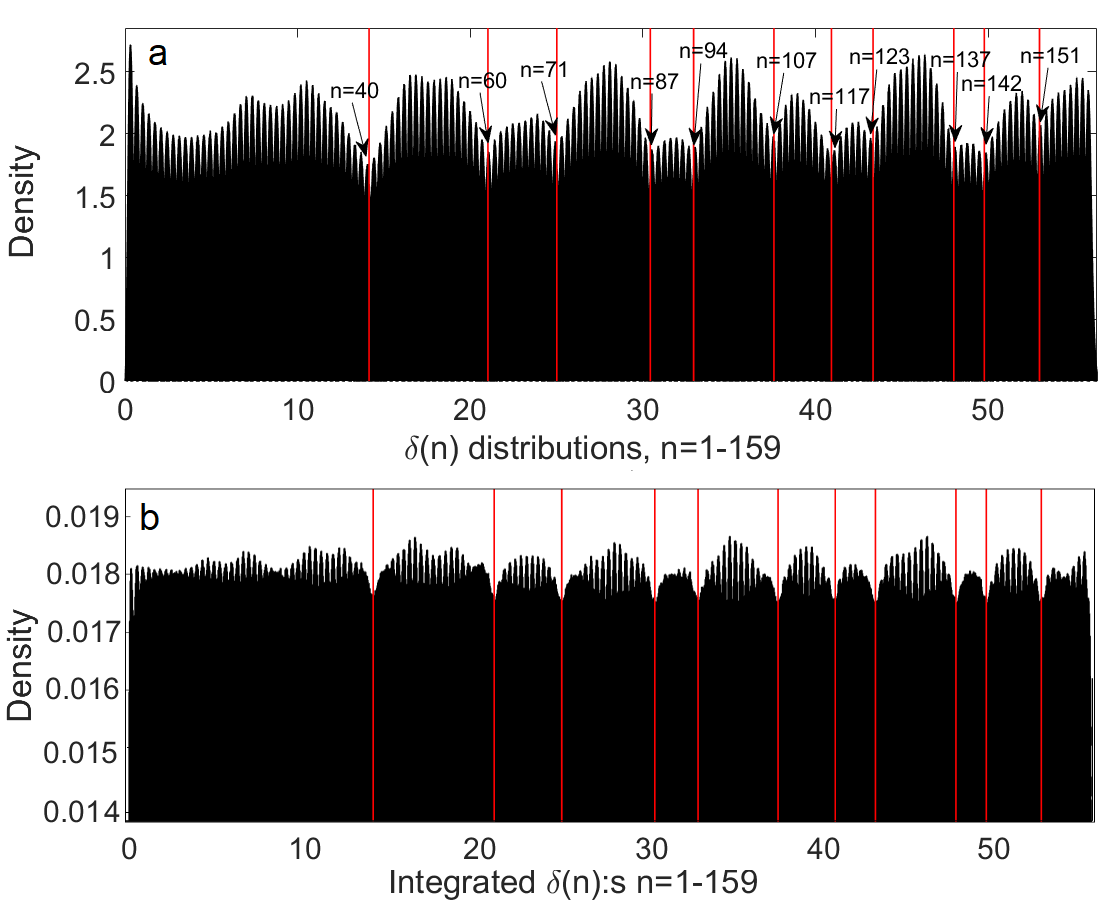}
		\caption{a) The distributions from the the height of one billionth zeta zero using $\delta(n)$ for n=1,2,3,..,159 . Note that the sites of first zeta zeros (marked with red vertical dashed line) are located near distributions with n=40, 60, 71, 87, 94, 107, 117, 123, 137, 142 and 151. b) The integrated distribution of the $\delta$-distributions of figure a.}
		\label{fig:distributions_1_milj}
\end{figure}

Here $i$ goes from j to j+5000000 in our analyses, where j is the ordinal number of the starting zeta zero. Because the zeros are not stabilized yet at the height of millionth zero we study, in the next, zeros and their differences at the height of 1, 10, 100 billionth zeta zero, and zeta zeros starting at $10^{23}$ \cite{Takalo}. Figure \ref{fig:distributions_1_milj}a shows the distributions of five million $\delta$s for n=1,2,3,..,159 from the the height of one billionth zeta zero. The distributions with n=40, 60, 71, 87, 94, 107, 117, 123, 137, 142 and 151 are the nearest distributions to the sites of first zeta zeros (marked with red vertical dashed line).  Figure \ref{fig:distributions_1_milj}b depicts the integrated distribution of the $\delta$-distributions in Fig. \ref{fig:distributions_1_milj}a. Note that in Fig. \ref{fig:distributions_1_milj}a each distribution is separately normalized to unit area, while for Fig. \ref{fig:distributions_1_milj}b of integrated distribution the whole area is one unit (we have cut part of the distribution to better show the troughs). We note that the troughs are clearer in Fig. \ref{fig:distributions_1_milj}a, where the distributions are the lower the nearer they are to the zeta zeros. When the distributions are summed together the troughs are partly filled with the tails of nearby distributions.

\begin{figure}
 \includegraphics[width=1.0\textwidth]{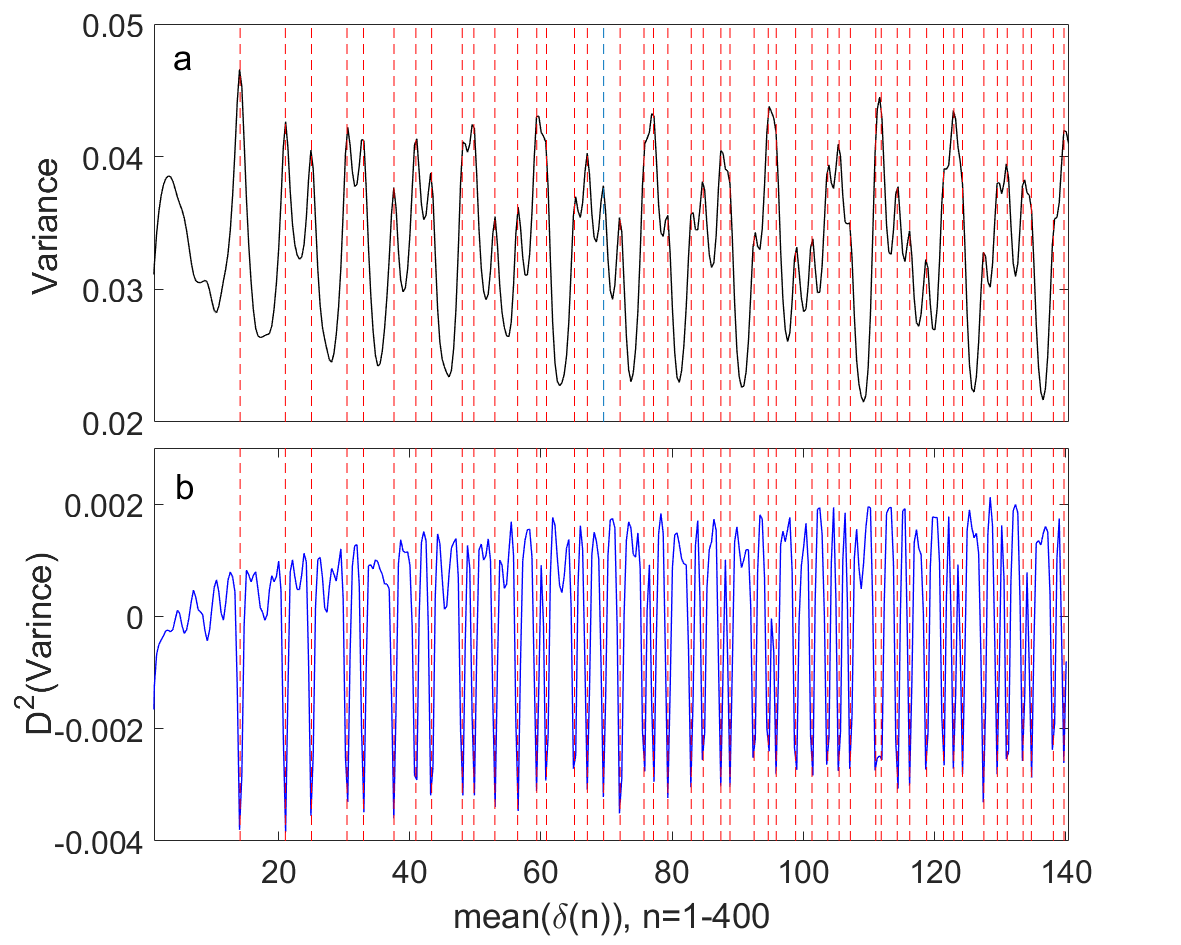}
\caption{a) The variances of the distributions of $\delta(n)$, n=1-400 from billionth zero.b) Second (discrete) derivative of the variance function of figure a.}
\label{fig:Variances_1_billion_1_400}     
\end{figure}

The distributions near the zeros have also characteristic features. Figure \ref{fig:Variances_1_billion_1_400}a depicts the variances of the distributions shown in Fig. \ref{fig:distributions_1_milj}, and Fig. \ref{fig:Variances_1_billion_1_400}b second (discrete) derivative of the variance function (we calculated these values a little further, i.e. for n=1-400). It is evident that variances have local maxima (or at least a turning point) at zeta zero. This is more clearly seen in Fig. \ref{fig:Variances_1_billion_1_400}b as a minimum of second derivative at each zeta zero. It, indeed, seems that the first zeros are encoded to the zeros (or their differences) at higher level.

\begin{figure}
 \includegraphics[width=1.0\textwidth]{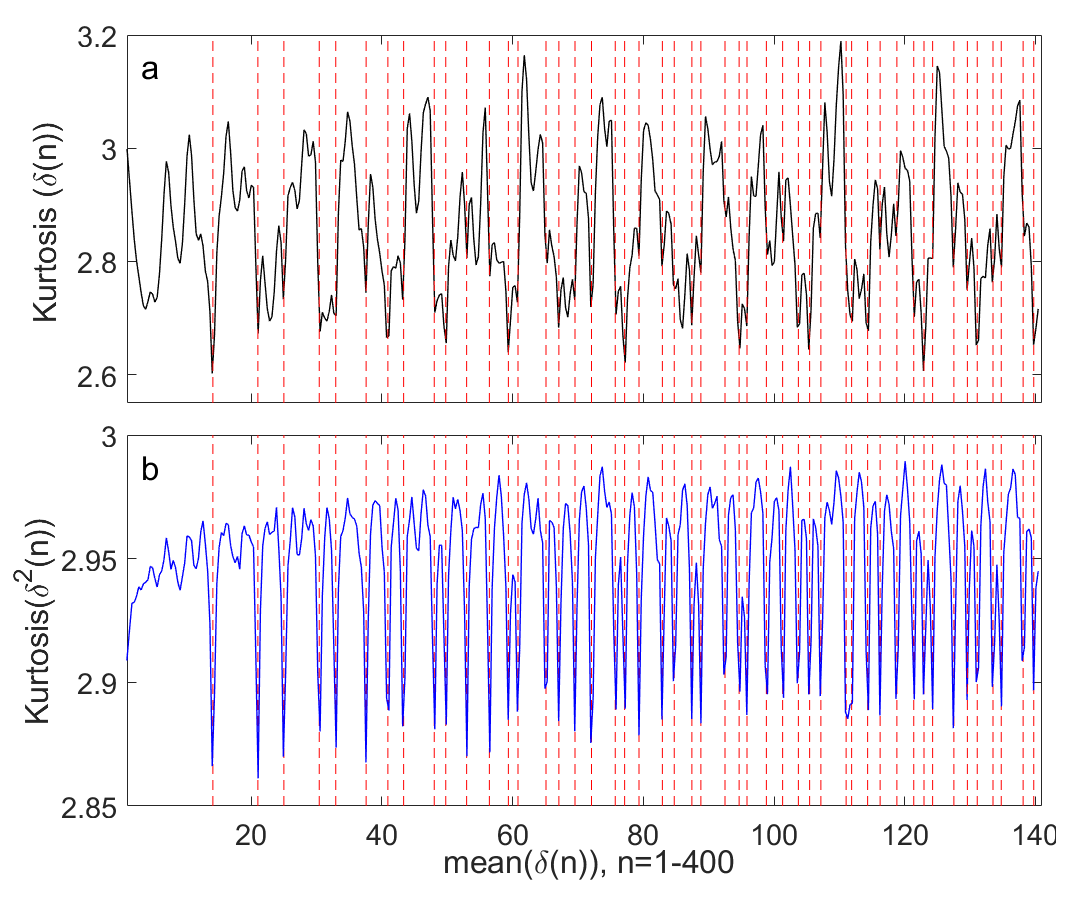}
\caption{a) The kurtoses of the distributions of $\delta(n)$, n=1-400 from billionth zero.b) The kurtoses of the distributions of $\delta^{2}(n)$, n=1-400 from billionth zero.}
\label{fig:Kurtoses_1_billion_1_400}     
\end{figure}

Figure \ref{fig:Kurtoses_1_billion_1_400}a depicts the kurtoses of the distributions for the same interval as Fig. \ref{fig:Variances_1_billion_1_400}. We note that the kurtoses have minima at (or near) every zeta zero. These are, however, not so clear than the maxima of the variances in Fig. \ref{fig:Variances_1_billion_1_400}a. Let us define the second difference, i.e. difference of the first difference $\delta(i)$ such that

\begin{equation}
\label{eq:delta2}
\delta^{2} = \delta(i+1)-\delta(i), i=1,2,3,...5000000-1
\end{equation}

If we plot the kurtoses for $\delta^{2}(i+n)-\delta^{2}(i)$ instead, the minima are much clearer, as seen in the Fig. \ref{fig:Kurtoses_1_billion_1_400}b. Now the local minima, which are not sites of zeta zeros, are tiny and clearly separable from the minima at the sites of zeta zeros. Note also the similarity of Figs. \ref{fig:Variances_1_billion_1_400}b and \ref{fig:Kurtoses_1_billion_1_400}b, only the vertical scale is different.  Again the information of the lower zeros is somehow contained in the differences of higher level zeros.

\begin{figure}
 \includegraphics[width=1.0\textwidth]{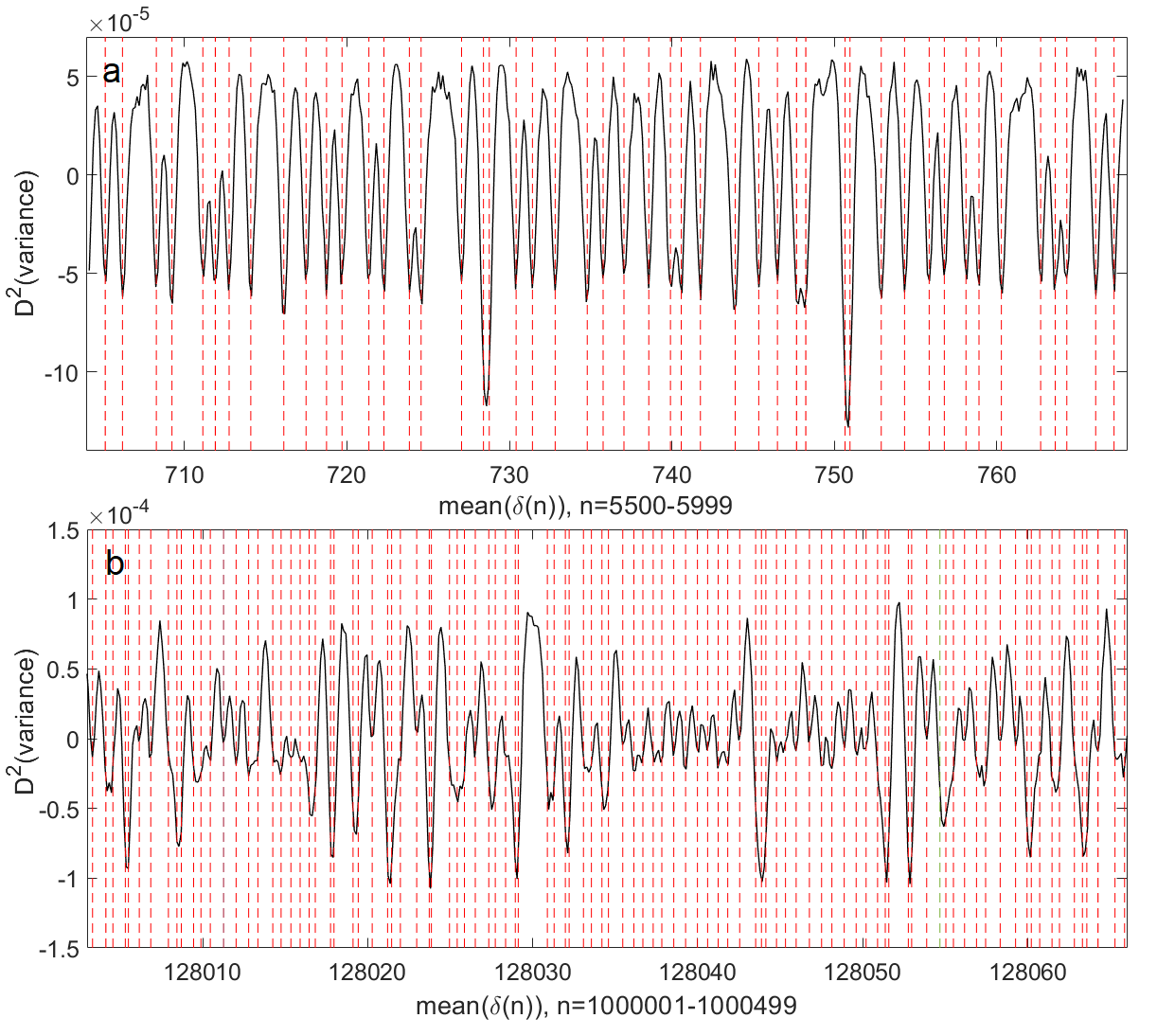}
\caption{a) Second derivative of the variances of the $\delta(n)$-distributions for n=5500-5999 calculated for 5 million zeta zeros starting from zero $10^{23}$ as a function of mean($\delta(n)$).b) Same as a, but for n=1000001-1000499. Zeta zeros are shown with vertical dashed red lines.}
\label{fig:Derivative_of_variances}     
\end{figure}

Figure \ref{fig:Derivative_of_variances} shows that the awareness of the other zeros is not restricted only to very first zeros. The variances of this figure are calculated from 5 million differences starting from $10^{23}$rd zero. The upper panel of Fig. \ref{fig:Derivative_of_variances} depicts the second (discrete) derivative of $\delta$(n)s with n=5500-5999. These distributions are located at height of the zeros between 504-568, and the zeros in this interval are shown as vertical dashed lines. The lower panel of Fig. \ref{fig:Derivative_of_variances} depicts the second derivative of $\delta$(n)s with n=1000001-1000499. These distributions are located at height of the zeros between 128002-128066. Notice that the intervals are equally long, although the latter has twice as many zeros as the first interval. The minima of the derivatives of the variances coincide quite well with the zeros at those intervals. However, the resolution is too poor to distinguish the two, very nearby, pairs. e.g. (728.405, 728.759) and (750.656, 750.966) in the upper panel or the triple peak (128043.51, 128043.85, 128044.12) in the lower panel as separate peaks \cite{Takalo}.

\subsection{Skewnesses of the $\delta$-distributions}

\begin{figure}
 \includegraphics[width=1.0\textwidth]{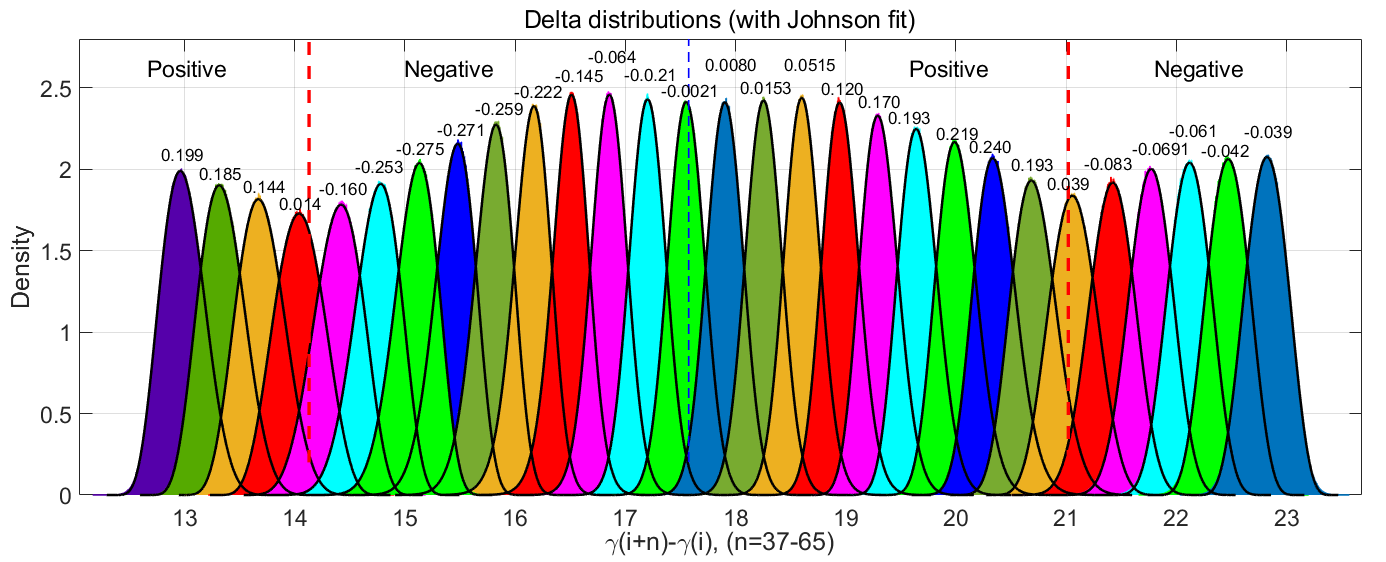}
\caption{The $\delta(n)$-distributions at height of billionth zeta zero using n=37-65 and fitted with Johnson probability density function. The decimal numbers are skewnesses of the separate distributions and the text tells their signs in each region.}
\label{fig:More_detailed_1_billion_deltas_37_65}     
\end{figure}

\begin{figure}
 \includegraphics[width=1.0\textwidth]{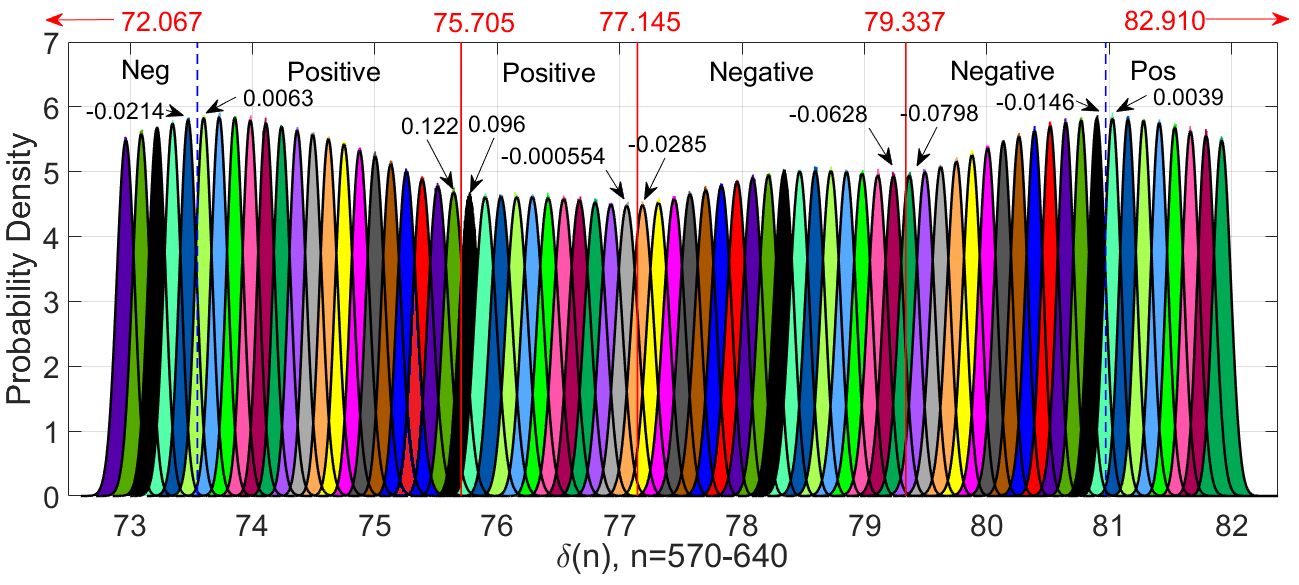}
\caption{The $\delta(n)$-distributions calculated using 5 million zeta zeros with n=280-350 starting from $10^{23}$rd zero. Distributions are fitted with Johnson distributions.}
\label{fig:Detailed_23_deltas_570_640}     
\end{figure}

Figure \ref{fig:More_detailed_1_billion_deltas_37_65} shows more detailed pattern of the $\delta$-distributions of Fig. \ref{fig:distributions_1_milj}a for n=37-65. The Johnson distribution fits are also shown in the figure. The red dotted vertical lines show the two first zeta zeros, and blue thinner dotted line locates in the halfway between these zeros. The decimal numbers tell the skewnesses of each distribution, and their sign is also shown as text above them. Note that the skewness changes sign at the zeros, and furthermore, at the halfway between the zeros such that $\delta$-distribution is here always skewed towards the nearest zeta zero. It seems that the $\delta$s of the zeros are avoiding the zeros themselves, i.e., Riemann zeros repel their $\delta$s \cite{Perez-Marco}.
Figure \ref{fig:Detailed_23_deltas_570_640} shows  the interval of $\delta$-distribution for n=570-640 starting from $10^{23}$rd zero. The situation is now more complicated, because the zeros are nearer to each other. In spite of this, the distributions are still lower at or near the sites of zeta zeros. The distributions in the left start with negative value for skewness, because there is a nearby zeta zero at 72.067 (not seen in the figure). The skewness changes from negative to positive at the blue dashed line, because of the next zero located at 75.705. After the zero the skewness, however, does not change to negative, but only decreases somewhat. It is evident that the two next zeros at right side 77.145 and 79.337 repel together more strongly than the only zero on the left side. The skewness changes sign then at 77.145, except that the last distribution at the left side is already slightly negative. Note also, that the skewness stays negative while passing the zero at 79.337, and only decreases again somewhat. This lasts until the blue dashed line, after which the skewness changes to positive due to the lurking zero at 82.910.

\subsection{Mean of the $\delta$-distributions}

\begin{figure}
 \includegraphics[width=0.85\textwidth]{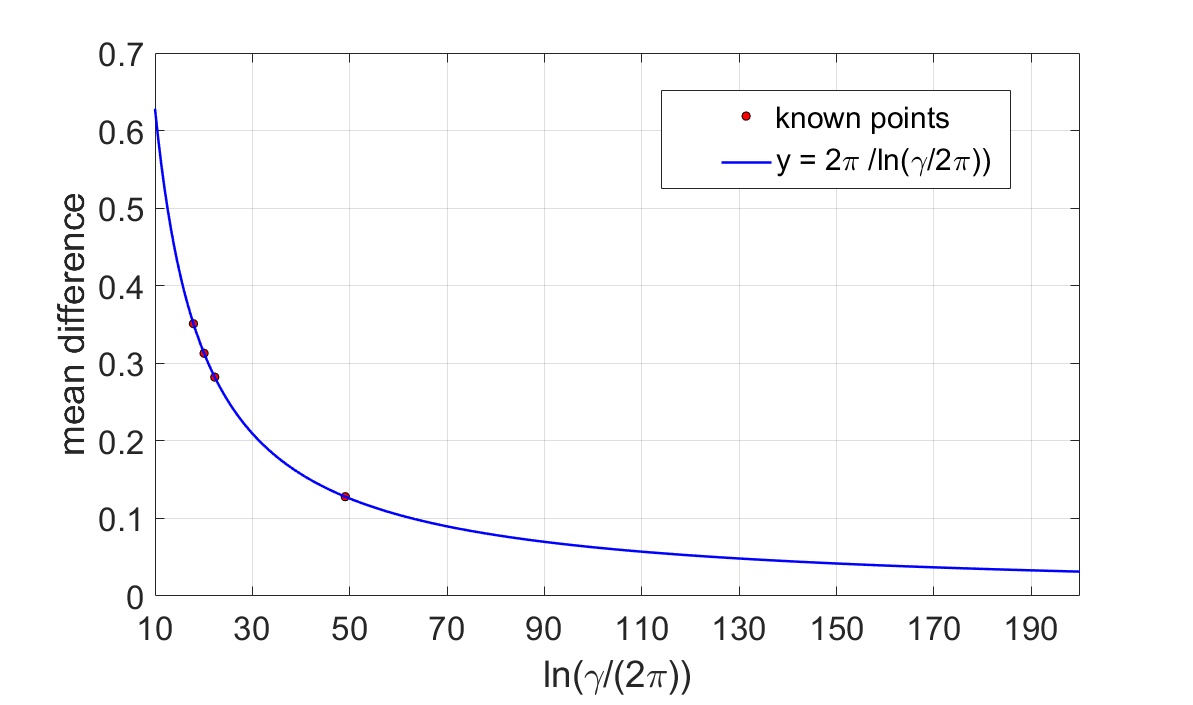}
\caption{A fit of the function $f(t)= (2\pi)/\ln(T/(2\pi))$ to the mean differences of zeta zeros; 0.351, 0.313, 0.282 and 0.128 for intervals at 1, 10, 100 billionth and $10^{23}$rd zeta zero, respectively.}
\label{fig:Mean_differences}     
\end{figure}

Figure \ref{fig:Mean_differences} shows a fit of the function $f(t)= (2\pi)/\ln(T/(2\pi))$ to the mean differences of zeta zeros; 0.351, 0.313, 0.282 and 0.128 for intervals of zeta zeros starting at 1, 10, 100 billionth and $10^{23}$rd zero, respectively. Note that the function is the inverted normalizing function of the first differences of the zeta zeros \cite{Odlyzko}. It is also clear that the mean difference of unscaled zeta zeros approaches zero, when the height of the zeros increase without bounds (at least for zeros on the line $\Re(z)=1/2$).

\begin{figure}
 \includegraphics[width=1.0\textwidth]{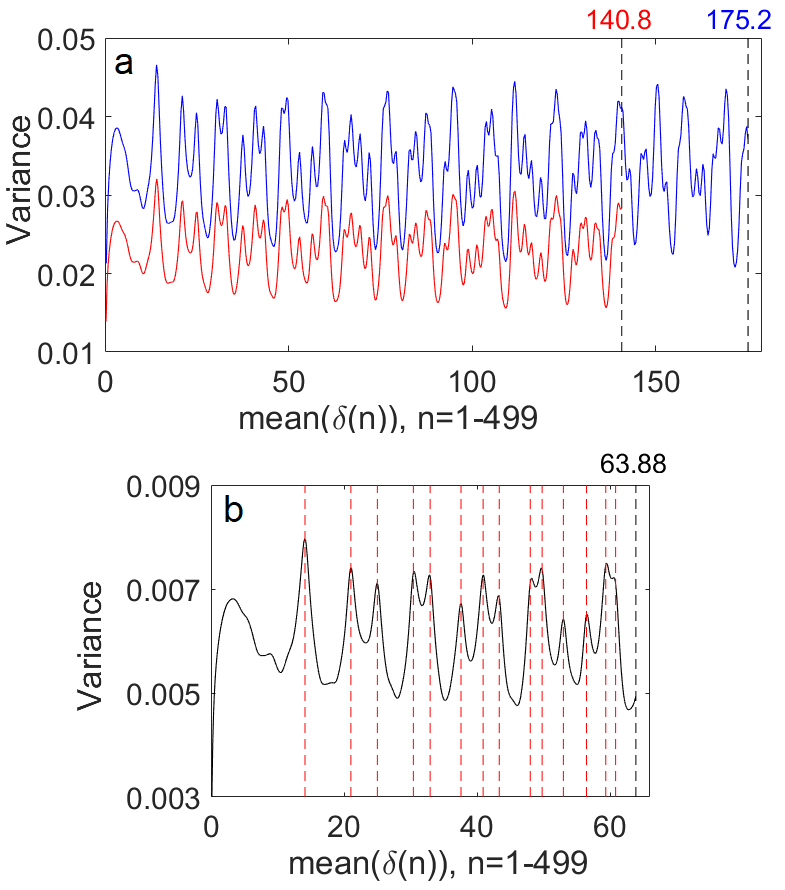}
\caption{a) Variances of $\delta$-distributions at 1 and 100 billionth zeta zeros, and b) of $\delta$-distributions starting at $10^{23}$rd zero for n=1-499 as a function of corresponding mean($\delta(n))$.}
\label{fig:Variances_1_499}     
\end{figure}

Figure \ref{fig:Variances_1_499}a shows variances for $\delta$-distributions at 1 and 100 billionth zeta zeros, and Fig. \ref{fig:Variances_1_499}b $\delta$-distributions starting at $10^{23}$rd zero for n=1-499 as a function of corresponding mean($\delta(n))$. The patterns of the variances are similar, except that their total length changes such that the same amount of distributions makes shorter variance curve when going higher level of the zeros. We plotted the first zeros (red vertical dashed lines) only to the Fig. \ref{fig:Variances_1_499}b for clarity. The last distributions seen in the variance plots are at mean value of $\delta$(499), i.e. at values 63.873, 140.779 and 175.171 for $10^{23}$rd, 100 billionth and billionth zeta zeros, respectively. The ratios of these numbers are 1:2.204:2.743. These numbers are inversely proportional to $ln(\gamma/(2\pi))$, where $\gamma$ is the value of average zeta zero in the corresponding zero interval (because the zeros are at so height level, we can here use the first zeros of the interval). The first zeros are $\gamma_{1}$=371870203.837,$\gamma_{2}$=29538618431.613 and $\gamma_{3}$=13066434408794275940027.301 for billionth, 100 billionth and $10^{23}$rd zeta zero. The ratios $1/ln(\gamma_{3}/(2\pi))$ : $1/ln(\gamma_{2}/(2\pi))$ : $1/ln(\gamma_{1}/(2\pi))$ are 1:2.204:2.743, which are same as aforementioned ratios of the lengths of the patterns of distributions. We could say that the higher we go in the zeros the less information the same amount of zeros contains about the lower zeros; the information content of the zeros at height $\gamma$ is inversely proportional to $ln(\gamma/(2\pi))$. However, while the mean difference of the zeros decreases also as inversely proportional to $ln(\gamma/(2\pi))$, this means that equally long segments of the line $\Re(z)=1/2$ contain equal amount of information, regardless of the height on the line.

\subsection{Skewness-kurtosis -plane}

\begin{figure}
 \includegraphics[width=1.0\textwidth]{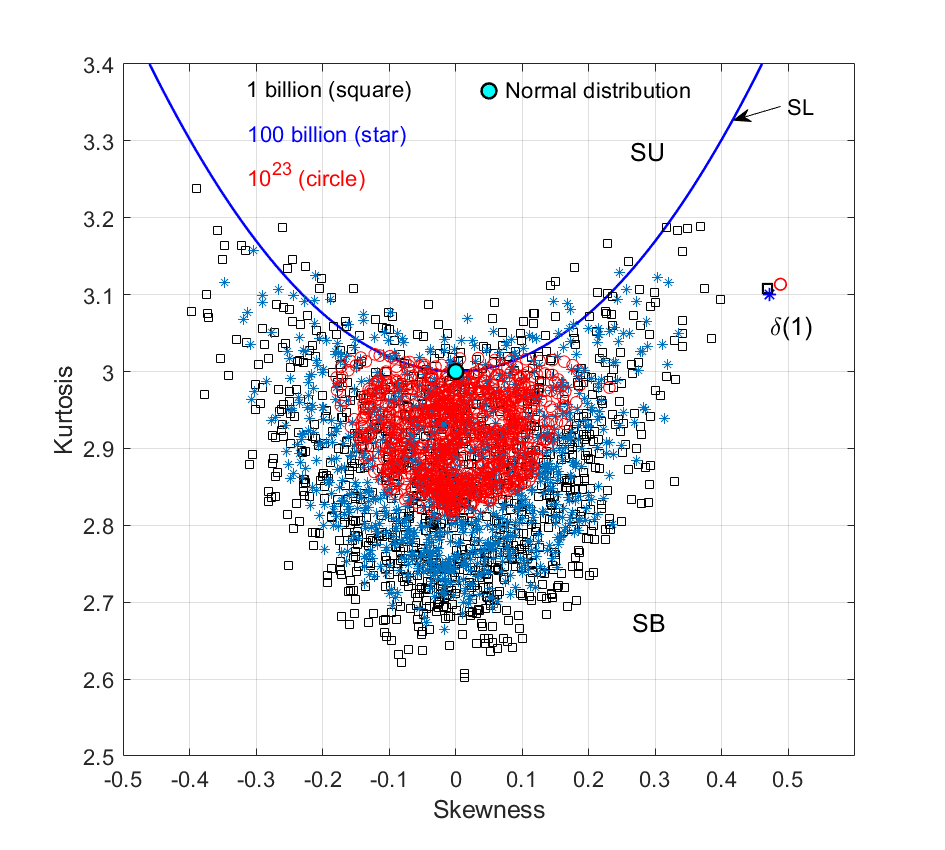}
\caption{The points of distributions starting from 1 billionth (black square), 100 billionth (light blue star) and $10^{23}$rd (red circle) zero in the Skewness-kurtosis -plane. The blue curve is Johnson SL distribution, which divides the plane to Johnson SU and SB regions. Cyan dot at (0,3) is the site of normal distribution in the skewness-kurtosis -plane.}
\label{fig:Skewness_kurtosis_plane}     
\end{figure}

As shown earlier the $\delta$-distributions can be fitted very well with the Johnson SB and SU PDFs (very rarely SL is needed). Figure  \ref{fig:Skewness_kurtosis_plane} shows the distributions in the skewness-kurtosis -plane \cite{Cugerone} for zeros at 1 billion, 100 billion and $10^{23}$. The red curve in the figure shows the border between Johnson SU and SB distributions (this is also region of SL distribution). It is notable that $\delta$(1)s are almost in the same place in the plane for all groups. Otherwise, the points seem to be more compactly located when going to higher levels of zeta zeros. Note, that all groups of distributions with n=2-999 form a heart-shaped pattern such that the point (0,3), which is the site of normal distribution in the skewness-kurtosis -plane, is located in the trough of the heart. We suppose that the average kurtosis is approaching the value 3, when going still to upper levels of zeta zeros, and the average skewness approaches zero from the positive side, i.e. distributions approach to normal distribution \cite{Takalo}. 
Furthermore, when going to the higher zeros the $\delta$-distributions are thinner and thinner, i.e, the standard deviation decreases. (We find that the standard deviation decreases somewhat slower than the mean value, i.e. proportional to about $2.1/(ln(\gamma/(2\pi)))^{0.843})$. We can then approximate the distributions (at very high levels of zeta zeros) with a normal distribution whose standard deviation approaches to zero, i.e. it tends to Dirac delta function at the mean value of the distribution. In this case we need more and more very nearby distributions to extract the information about the earlier zeros of the zeta function. 

\section{Prime numbers and Riemann zeta zero spectrum}

\begin{figure}
 \includegraphics[width=1.0\textwidth]{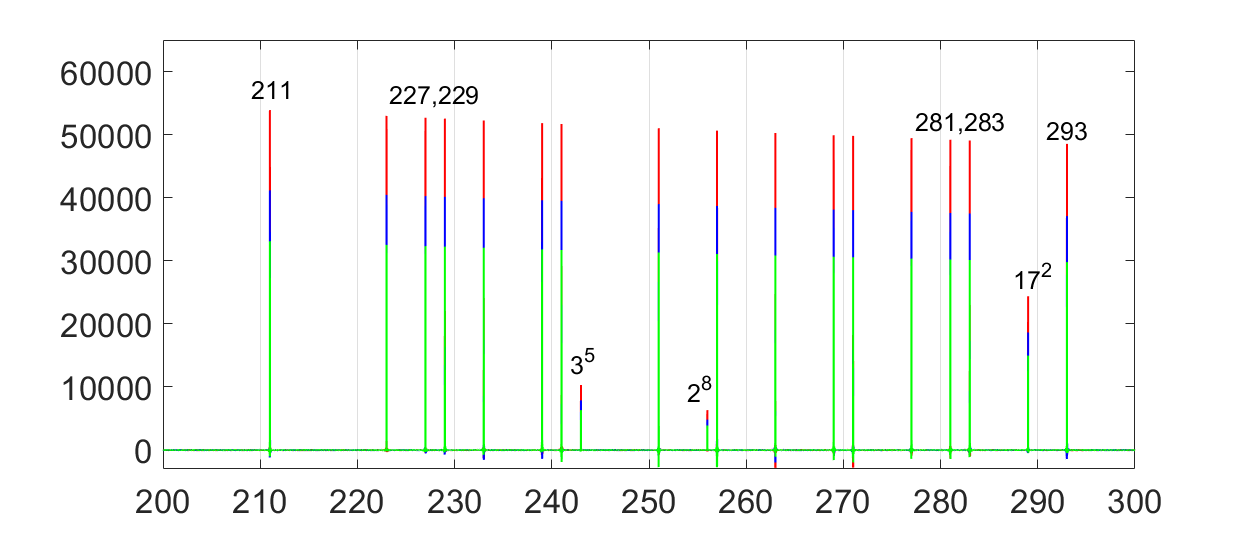}
\caption{The primes and their powers between 200-300 calculated from Eq. \ref{eq:primes} using 2 million zeta zeros starting at ten millionth (red curve), billionth (blue curve) and 100 billionth zeta zero (green curve).}
\label{fig:primes_200_300}     
\end{figure}

We calculate the function of prime powers from the cosine series \cite{Mazur}

\begin{equation}
\label{eq:primes}
	P(x) = -\sum^{N}_{i=1}cos\left(\gamma_{i}\,\times\,ln(x)\right) ,
\end{equation}
where $\lambda_{i}$ is the first element in the sequence of the zeta zeros and N is the number of zeros in the sequence.
Figure \ref{fig:primes_200_300} depicts the primes and their powers between 200-300 calculated from \ref{eq:primes} using 2 million zeta zeros starting at 10 millionth (red curve), billionth (blue curve) and 100 billionth zeta zero (green curve). There are four twin primes (227/229, 239/241, 269/271 and 281/283), three prime powers ($3^{5}$, $2^{8}$ and $17^{2}$) and eight single primes in this interval. The interesting thing here is that the two million zeros at height 10 millionth zero give stronger peaks at the sites of the primes than same amount of zeros at height 1 billionth zero, which in turn give stronger peaks than two million zeros at height 100 billionth zero. The ratios of the heights are 1:1.244:1.629, which is same ratio as $1/ln(\gamma_{1}/(2\pi))$ : $1/ln(\gamma_{2}/(2\pi))$ : $1/ln(\gamma_{3}/(2\pi))$, where $\gamma_{1}$, $\gamma_{2}$ and $\gamma_{3}$ are the average zeros at height 100 billion, billion and 10 million, respectively, i.e, the strength of the line is (again) inversely proportional to ln($\gamma$/(2$\pi$).
Note that the peaks of prime powers are lower than the other prime peaks. This is because peaks are scaled with the power such that the strength of the peak decreases to $1/p^{(n-1)/2}$, where n=2,3,4,...is the power of the original prime p.

\begin{figure}
 \includegraphics[width=1.0\textwidth]{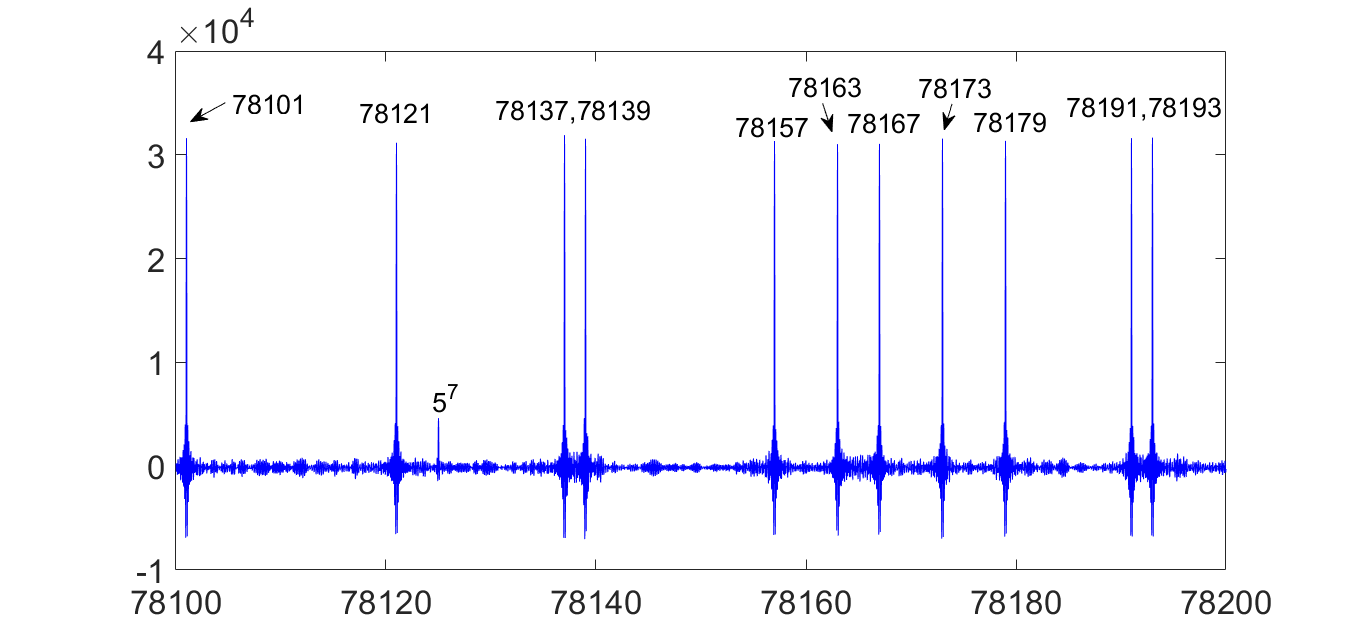}
\caption{The primes and prime power $5^{7}$ between 78100-78200 calculated from first ten million zeta zeros.}
\label{fig:primes_at_78100}     
\end{figure}

We note also that the peaks are decreasing when going further in the real axis. This is probably due to the logarithmic dependence of the x (see Eq. \ref{eq:primes}) such that while period in constantly increasing the amplitude decreases and eventually dies out \cite{Sakhr}.
Also the extra fluctuation around (and between) the lines of primes increases when going upper in the real axis. Figure \ref{fig:primes_at_78100} depicts the primes between 78100-78200 calculated from first ten million zeta zeros. There are two twin primes, seven single primes and one prime power ($5^{7}$) between this interval. Although we used so many zeta zeros the result is not anymore satisfying. In order to plot large primes we need a huge amount of zeros and capacity in calculation. The extra fluctuation can be diminished using windowing, but anyhow this method is impractical for finding new primes \cite{Sakhr}.

\section{Conclusion}

We have studied the $\delta$-distributions of zeros of Riemann zeta function at heights 10 millionth, billionth 100 billionth and $10^{23}$ -rd zero such that we calculate distribution for each difference, $\delta(n)$, separately. We used 5 million $\delta$s for these analyses, and showed that statistical properties are very similar for all intervals. It is interesting that a finite subset of successive zeros, independently of its height, from the infinite sequence of nontrivial zeros knows the location of other zeros. The information contained in the subset of zeros in inversely proportional to $ln(\gamma/(2\pi))$, where $\gamma$ is the average of the zeros in the subset. However, while the density of the zeros increases as proportional to $ln(\gamma/(2\pi))$, each equally long  segment of the line $\Re(z)=1/2$ contains equal amount of information. The same proportionality $ln(\gamma/(2\pi))$ exists also in the strength of the lines of the prime powers calculated from subset of zeta zeros.

The skewness of the $\delta$-distributions change sign, when crossing zeta zero or, at least, decreases when passing the zero in increasing direction. The variance has local maximum or turning-point and the kurtosis local minimum at zeta zero.
We also plotted $\delta(n)$-distributions (for n=1-999) of zeta zeros at heights 1 billion, 100 billion and $10^{23}$ on the skewness-kurtosis -plane. All these groups of points form a heart-shaped pattern such that the higher the zeros the more compactly the corresponding points are located in the skewness-kurtosis -plane. The site of normal distribution in the plane, i.e. point (0,3) is in the trough of this pattern. We believe, that going still higher levels in the zeros the average kurtosis approaches value 3, which is the kurtosis of normal distribution. The $\delta$(1) is located almost at the same point for all groups apart from other points of the corresponding group.

\begin{acknowledgements}
We acknowledge LMFDB and A.M. Odlyzko for the zeta zero data.
\end{acknowledgements}

%
\section*{Conflict of interest}
The authors declare that they have no conflict of interest.


\end{document}